\documentclass[twocolumn,aps,showpacs,floatfix,prc]{revtex4}
\usepackage[dvips]{epsfig}
\begin{document}
%\draft

\title{ Fusion cross sections for reactions involving medium $\&$ heavy nucleus-nucleus systems }
\author{ Debasis Atta$^1$ and D.N. Basu$^2$ }

\affiliation{ Variable  Energy  Cyclotron  Centre, 1/AF Bidhan Nagar, Kolkata 700 064, India }

\email[E-mail 1: ]{datta@vecc.gov.in}
\email[E-mail 2: ]{dnb@vecc.gov.in}

\date{\today }

\begin{abstract}

    Existing data on near-barrier fusion excitation functions of medium and heavy nucleus-nucleus systems
have been analyzed using a simple diffused barrier formula derived assuming the Gaussian shape of the
barrier height distributions. Fusion cross section is obtained by folding the Gaussian barrier distribution with the classical expression for the fusion cross section for a fixed barrier. The energy dependence of the fusion cross section, thus obtained, provides good description to the existing data on near-barrier fusion and capture excitation functions for medium and heavy nucleus-nucleus systems. The fusion or capture cross section predictions are especially important for planning experiments for synthesizing new super-heavy elements.

\vskip 0.2cm
\noindent
{\it Keywords}: Fusion reactions; Barrier distribution; Excitation function; Synthesis of elements.  
\end{abstract}

\pacs{ 25.70.-z; 25.60.Pj; 26.20.Np; 97.10.Cv }   
\maketitle

\noindent
\section{Introduction}
\label{section1}

    Nuclear fusion reactions are widely used in nuclear physics to produce nuclei far from the $\beta$- stability line and superheavy nuclei, to explore the properties of excited nuclear states and the mechanisms of their decay and to study the dynamics of nuclear reactions \cite{Ho78,Ba80,No80,Sa83,Fe92,Fr96}. The burning of stars is also associated with reactions involving the sub-barrier fusion of nuclei \cite{Wa97}. Fusion of two atomic nuclei occurs when the interacting bodies can overcome the barrier formed by the sum of the attractive nuclear and the repulsive Coulomb and centrifugal potentials. If the kinetic energy measured in the center of mass system is below the barrier height, classically forbidden fusion can occur through quantum tunneling of the barrier which is known as sub-barrier fusion. The nuclear fusion cross section at energies in the vicinity of the barrier depends greatly on the barrier height. It is well known that fusion excitation functions cannot be satisfactorily explained assuming penetration through a single, well-defined barrier in the total potential energy of a colliding nucleus-nucleus system. In order to reproduce shapes of the fusion excitation functions, especially at low near-threshold energies, it is necessary to assume coexistence of different barriers, a situation that is naturally accounted for in the description of fusion reactions in terms of coupled channel calculations involving coupling to various collective states.

    The aim of the present work is to obtain the nuclear fusion cross sections for reactions involving medium and heavy nucleus-nucleus systems. The phenomenological description of the fusion excitation functions is achieved by assuming a Gaussian shape of the barrier distribution treating the mean barrier and its variance as free parameters and folding it with the classical expression for the fusion cross section for a fixed barrier with the distance corresponding to the location of the interaction barrier as another free parameter. The free parameters are then determined individually for each of the reactions by comparing the predicted fusion excitation function with experimental data. The energy dependence of the fusion cross section, thus obtained, provides good description to the existing data on near-barrier fusion and capture excitation functions for medium and heavy nucleus-nucleus systems. The predictions of fusion or capture cross sections are especially important for planning experiments aimed at producing new super-heavy elements.

\noindent
\section{Fusion barrier distribution}
\label{section2}

    It is well known that the energy dependence of the fusion cross sections can not be well estimated assuming simply the penetration through a well-defined barrier in one-dimensional potential of a colliding nucleus-nucleus system. The heavy-ion fusion cross sections require interpretation \cite{Ro91} in terms of a distribution of potential barriers. The smoothening due to the quantal barrier penetration replaces set of discrete barriers by an effective continuous distribution. In order to reproduce shapes of experimentally observed fusion excitation functions, particularly at low, near-threshold energies, it is necessary to assume a distribution of the fusion barrier heights, the effect that results from the coupling to other than relative distance degrees of freedom. This is naturally achieved in coupled-channel calculations, involving the coupling to the lowest collective states in both colliding nuclei. The structure effects in the barrier distributions are neglected in the present work and for the distribution of the fusion barrier heights, a Gaussian shape for the barrier distribution $D(B)$ is assumed \cite{Wi04}. The barrier distribution is, therefore, given by

\begin{equation}
 D(B)dB=\frac{1}{\sqrt{2\pi}\sigma_B}\exp\Big[-\frac{(B-B_0)^2}{2\sigma_B^2}\Big] 
\label{seqn1}
\end{equation}
\noindent
where the two parameters, the mean barrier $B_0$ and the distribution width $\sigma_B$, to be determined individually for each reaction.

\noindent
\section{The fusion cross section}
\label{section3}

    The energy dependence of the fusion cross section is obtained by folding the barrier distribution \cite{Wi04,Ca11} provided by Eq.(1), with the classical expression for the fusion cross section given by

\begin{eqnarray}
 \sigma_{f}(B) =&& \pi R_B^2 \Big[1-\frac{B}{E}\Big] ~~~~~~~~~~~~~~{\rm for} ~~B\leq E  \nonumber\\
 =&&0 ~~~~~~~~~~~~~~~~~~~~~~~~~~~~~~{\rm for}~~B\geq E
\label{seqn2}
\end{eqnarray}
\noindent
where $R_B$ denotes the relative distance corresponding to the position of the barrier approximately, which yields

\begin{eqnarray}
 &&\sigma_{fus}(E) = \int_0^\infty \sigma_{f}(B) D(B)dB \nonumber\\
 &&= \int_0^{B_0} \sigma_{f}(B) D(B)dB + \int_{B_0}^E \sigma_{f}(B) D(B)dB  \\
 &&= \pi R_B^2\frac{\sigma_B}{E\sqrt{2\pi}}\Big[\xi\sqrt{\pi}\Big\{{\rm erf}\xi+{\rm erf}\xi_0\Big\}   
 +e^{-\xi^2} +e^{-\xi_0^2}\Big]  \nonumber
\label{seqn3}
\end{eqnarray}
\noindent
where

\begin{eqnarray}
 \xi &&= \frac{E-B_0}{\sigma_B\sqrt{2}} \nonumber\\
 \xi_0 &&= \frac{B_0}{\sigma_B\sqrt{2}} 
\label{seqn4}
\end{eqnarray}
\noindent
and erf$\xi$ is the Gaussian error integral of argument $\xi$. The parameters $B_0$ and $\sigma_B$ along with $R_B$ is to be determined by fitting Eq.(3) along with Eq.(4) to a given fusion excitation function. In the derivation of formula Eq.(3), the quantum effect of sub-barrier tunneling is not accounted for explicitly. However, the influence of the tunneling on shape of a given fusion excitation function is effectively included in the width parameter $\sigma_B$. 

    The fusion cross section formula of the Eq.(3) obtained by using the diffused-barrier, is a very elegant parametrization of the cross section for a process of overcoming the potential-energy barrier. Hence, it can be successfully used for analysis and predictions of the fusion excitation functions of light, medium and moderately heavy systems, especially in the range of near-barrier energies. The term `capture' is used to refer the process of overcoming the interaction barrier in a nucleus-nucleus collision, followed by formation of a composite system. In general, the composite system undergoes fusion only in a fraction $f$ of the capture events. For light and medium systems, $f\approx1$, and almost all the `capture' events lead to fusion resulting fusion cross sections to be practically identical with the capture cross sections. However, for very heavy systems, only a small fraction ($f<1$) of `capture' events ultimately lead to fusion while for the remaining part of the events, the system re-separates prior to equilibration and clear distinction between fusion and capture cross sections then becomes necessary. Therefore, for very heavy systems, when the overcoming the barrier does not guarantee fusion, predictions based on Eq.(3) give the capture cross section.

\noindent
\section{ Calculation and results }
\label{section4}
    
    The near-barrier (above barrier) fusion excitation functions of medium and heavy nucleus-nucleus systems
have been analyzed using a simple diffused barrier formula (given by Eq.(1)) derived by folding the Gaussian barrier distribution with the classical expression for the fusion cross section for a fixed barrier. The same set of target-projectile combinations have been selected for which heavy ion sub-barrier fusion has been recently \cite{Ro13} studied. The values of mean barrier height $B_0$, width $\sigma_B$ and the effective radius $R_B$ have been obtained using the least-square fit method. These values are listed in Table-I and arranged in order of the increasing value of the Coulomb parameter $z=Z_1Z_2/(A_1^{1/3} + A_2^{1/3})$. Since the number of data points for $^{48}$Ca+$^{124}$Sn is too low compared to other systems, the error bars for $B_0$ (111.93$\pm$0.44), $\sigma_B$ (1.28$\pm$0.83) and $R_B$ (8.24$\pm$0.09) are rather large.

\begin{table}[h]
\vspace{0.0cm}
\caption{\label{tab:table1} The values of the mean barrier height $B_0$, the width of the barrier height distribution $\sigma_B$ and the effective radius $R_B$, deduced from the analysis of the measured fusion excitation functions.}
\begin{tabular}{|c|c|c|c|c|c|}
\hline
Reaction & $z$ &Refs.&$B_0$&$\sigma_B$&$R_B$ \\
& & & [MeV] & [MeV] & [fm]         \\  
\hline

$^{16}$O+$^{154}$Sm& 62.94 & \cite{Le95} &58.80 &2.43 &10.04 \\
$^{17}$O+$^{144}$Sm& 63.49 & \cite{Le95} &60.28 &1.75 &10.46 \\
$^{16}$O+$^{148}$Sm& 63.51 & \cite{Le95} &59.88 &2.31 &10.61 \\ 
$^{16}$O+$^{144}$Sm& 63.91 & \cite{Le95} &60.65 &1.76 &10.46 \\
$^{36}$S+$^{110}$Pd& 90.94 & \cite{St95} &85.51 &1.92 &8.20 \\   
$^{32}$S+$^{110}$Pd& 92.39 & \cite{St95} &86.04 &3.10 &8.45 \\  
$^{48}$Ca+$^{96}$Zr& 97.41 & \cite{St06} &93.76 &2.75 &10.07 \\ 
$^{48}$Ca+$^{90}$Zr& 98.58 & \cite{St06} &94.94 &2.11 &10.01 \\
$^{40}$Ca+$^{96}$Zr& 100.01& \cite{Ti98} &94.30 &3.09 &9.71  \\ 
$^{40}$Ca+$^{90}$Zr& 101.25& \cite{Ti98} &96.26 &1.67 &10.07 \\ 
$^{48}$Ca+$^{124}$Sn&116.00& \cite{Ko12} &111.93&1.28 &8.24 \\ 
$^{40}$Ca+$^{124}$Sn&118.95& \cite{Sc00} &113.36 &2.70 &9.57 \\ 

\hline
\end{tabular} 
\vspace{0.0cm}
\end{table}

    In Figs.-1 $\&$ 2, the measured fusion excitation functions represented by full circles are compared with the predictions of the diffused barrier formula depicted by the solid lines. The two systems of $^{16}$O+$^{144}$Sm and $^{40}$Ca+$^{124}$Sn illustrated in Figs.-1 $\&$ 2 correspond to two extreme Coulomb parameter ($z$) values of $\sim$ 64 and $\sim$ 119, respectively. It can, therefore, be easily perceived from these figures that precisely measured fusion excitation functions provide systematic information on the essential characteristics of the interaction potential, {\it viz.} the mean barrier height $B_0$ and width $\sigma_B$ of its distribution, for nucleus-nucleus collisions. The fusion or capture cross sections can also be predicted by using Eq.(3) and theoretically obtained values of the parameters $B_0$ and $\sigma_B$ \cite{Wi04} for planning experiments for synthesizing new super-heavy elements.

\begin{figure}[t]
\vspace{0.8cm}
\eject\centerline{\epsfig{file=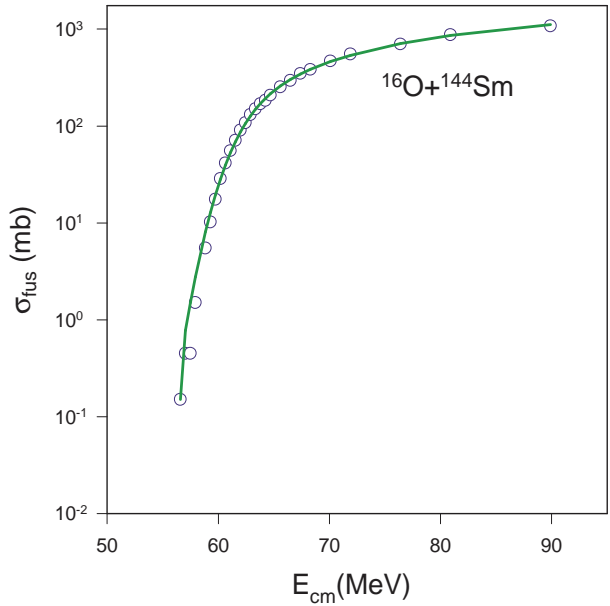,height=8cm,width=8cm}}
\caption
{Comparison of the measured fusion excitation functions (full circles) for $^{16}$O+$^{144}$Sm with predictions (solid lines) of the diffused barrier formula.}
\label{fig1}
\vspace{0.0cm}
\end{figure}
\noindent 

\begin{figure}[t]
\vspace{0.8cm}
\eject\centerline{\epsfig{file=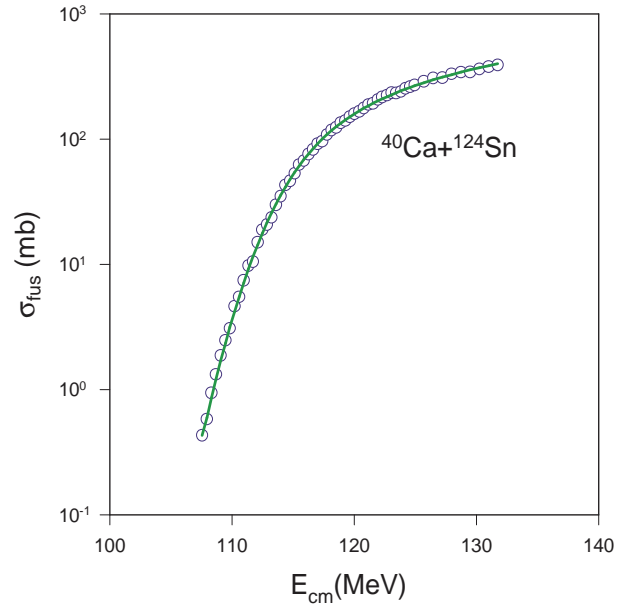,height=8cm,width=8cm}}
\caption
{Comparison of the measured fusion excitation functions (full circles) for $^{40}$Ca+$^{124}$Sn for with predictions (solid lines) of the diffused barrier formula.}
\label{fig2}
\vspace{0.0cm}
\end{figure}
\noindent 

\noindent
\section{ Summary and conclusion }
\label{section5}

    In summary, the fusion reaction cross sections have been calculated for above barrier energies over a wide energy range. A set of precisely measured fusion excitation functions has been studied in order to learn about the conditions of overcoming the potential energy barrier in nucleus-nucleus collisions and to obtain systematic information on the essential characteristics of the interaction potential, {\it viz.} the mean barrier height $B_0$ and width $\sigma_B$ of its distribution, between the two colliding nuclei. For the analysis of the experimental data a simple diffused-barrier formula is derived assuming Gaussian shape for the barrier distribution. Using the least-square fit method, precisely determined values of the mean barrier height $B_0$, the width $\sigma_B$ and the effective radius $R_B$ have been obtained.           

    Results of the present analyses provide a new vista to predict fusion excitation functions for not yet studied nuclear systems. Strictly speaking, one can predict only the `overcoming-the-barrier' cross sections which, as discussed are not identical with fusion cross sections in the case of very heavy systems. To calculate the overcoming-the-barrier cross section for a given projectile-target combination one can use the present fusion cross section formula and apply theoretical values of the parameters $B_0$ and $\sigma_B$ to predict cross sections for overcoming the barrier in collisions of very heavy systems used to produce superheavy nuclei. Prediction of the energy dependence of the cross section for capture or sticking can be used as one of three basic ingredients in the sticking-diffusion-survival model \cite{Sw03} for calculating the production cross sections of superheavy nuclei. \\

\end{document}